\documentclass[aps,pre,twocolumn,floatfix,superscriptaddress,showpacs,showkeys]{revtex4-1}
\usepackage{epsf,amsmath,amssymb,verbatim,color,multirow,pifont,graphicx,hyperref,cleveref,tabularx,xcolor}

\usepackage{soul}

\crefname{align}{Eq.}{Eqs.}
\crefname{subsection}{Sec.}{Secs.}
\crefname{section}{Sec.}{Secs.}
\crefname{figure}{Fig.}{Figs.}
\crefname{table}{Table}{Tables.}

\begin{document}

\title{Effective potential approach to hybrid synchronization transitions}
\author{Je Ung Song}
\affiliation{CCSS, CTP and Department of Physics and Astronomy, Seoul National University, Seoul 08826, Korea}
\author{Jaegon Um}
\affiliation{CCSS, CTP and Department of Physics and Astronomy, Seoul National University, Seoul 08826, Korea}
\affiliation{BK21PLUS Physics Division, Pohang University of Science and Technology, Pohang 37673, Korea}
\author{Jinha Park}
\affiliation{CCSS, CTP and Department of Physics and Astronomy, Seoul National University, Seoul 08826, Korea}
\author{B. Kahng}
\email{bkahng@snu.ac.kr}
\affiliation{CCSS, CTP and Department of Physics and Astronomy, Seoul National University, Seoul 08826, Korea}
\date{\today}

\begin{abstract}
The Kuramoto model exhibits different types of synchronization transitions depending on the type of natural frequency distribution. To obtain these results, the Kuramoto self-consistency equation (SCE) approach has been used successfully. However, this approach affords only limited understanding of more detailed properties such as the stability and finite size effect. Here, we extend the SCE approach by introducing an effective potential, that is, an integral version of the SCE. We examine the landscape of this effective potential for second-order, first-order, and hybrid synchronization transitions in the thermodynamic limit. In particular, for the hybrid transition, we find that the minimum of effective potential displays a plateau across the region in which the order parameter jumps. This result suggests that the effective free energy can be used to determine a type of synchronization transition. For finite systems, the effective potential contains local minima at which the system can be trapped. Using numerical simulations, we determine the stability of the system as a function of system size and simulation time.
\end{abstract}

\maketitle

\section{Introduction}
Phase transitions in equilibrium systems are conventionally classified according to the Ehrenfest classification scheme~\cite{ehrenfest}. When the $n$-th derivative of the free energy with respect to its argument first becomes discontinuous, the phase transition is of the $n$-th order. Many phase transitions are either second-order or first-order, where an order parameter such as the magnetization changes from zero to a finite value continuously or discontinuously, and fluctuations are divergent or finite, respectively. However, this classification scheme does not accommodate some phase transitions. For instance, although the order parameter is discontinuous, critical behavior appears at the same transition point; e.g., the fluctuations of the order parameter and/or the correlation length diverge. 
This type of abnormal phase transition is called a mixed-order transition. In addition, a new type of phase transition has been observed, in which the order parameter exhibits first-order and second-order transition behavior at the same transition point. This type of transition is called a hybrid phase transition (HPT). The terms mixed-order and HPT may often be used interchangeably. Examples appear in various equilibrium and nonequilibrium systems, including the Ising model with long-range interactions in one dimension~\cite{thouless_1969, mukamel, dyson, aizenman}, the Ashkin--Teller (AT) model on scale-free networks~\cite{at}, $k$-core percolation~\cite{kcore1, kcore2, kcore3, kcore4}, DNA denaturation~\cite{dna1, dna2, dna3}, jamming~\cite{jamming1, jamming2, jamming3}, crystallization of colloidal magnets~\cite{colloid}, and synchronization~\cite{sync_pazo, sync_moreno, sync_mendes}.  

Landau theory has been useful for determining the type of phase transition in equilibrium systems and determining the critical exponents in the mean-field limit for the second-order transition. The Landau free energy $\mathcal{L}(m)$ in Euclidean space is expanded with respect to the order parameter $m$ (the magnetization) in polynomial form as 
\begin{equation}
\mathcal{L}(m)=\frac{1}{2}(T-T_x)m^2-\frac{1}{3}a_3m^3+\frac{1}{4}a_4 m^4+\cdots .
\label{eq:landau}
\end{equation}
For the second-order transition, $a_3$ can be zero when $\mathcal L(m)=\mathcal L(-m)$ is symmetric, and $a_4 >0$. $T_x$ is a transition point $T_c$, across which the position of the global minimum of $\mathcal L(m)$ changes from $m=0$ for $T > T_c$ to finite $m$ (e.g., $m > 0$) for $T < T_c$. $\partial \mathcal{L}/\partial m=0$ and $\partial^2 \mathcal{L}/\partial m^2 < 0$ at $m=0$.  
For the first-order transition, $a_3 > 0$. $\mathcal L(m)$ has a minimum at $m=0$ for $T > T_x$. Moreover, there exists $m^*(T) > 0$ such that $\partial \mathcal{L}/\partial m =0$ at $m^*$ when $a_3^2 > 4a_4(T-T_x)$ for $T > T_x$. The local minimum of $\mathcal L$ at $m^*$ becomes a global minimum at $T_c$. Then, for $T < T_c$, a global free energy minimum exists at $m=m^*$. Thus, the first-order transition occurs at $T_c$, which is higher than $T_x$. Therefore, the order parameter is discontinuous across $T_c$. We remark that $\partial^2 \mathcal{L}/\partial m^2 > 0$ at $m=0$ for $T_c$. 

\begin{figure}
\includegraphics[width=.95\linewidth]{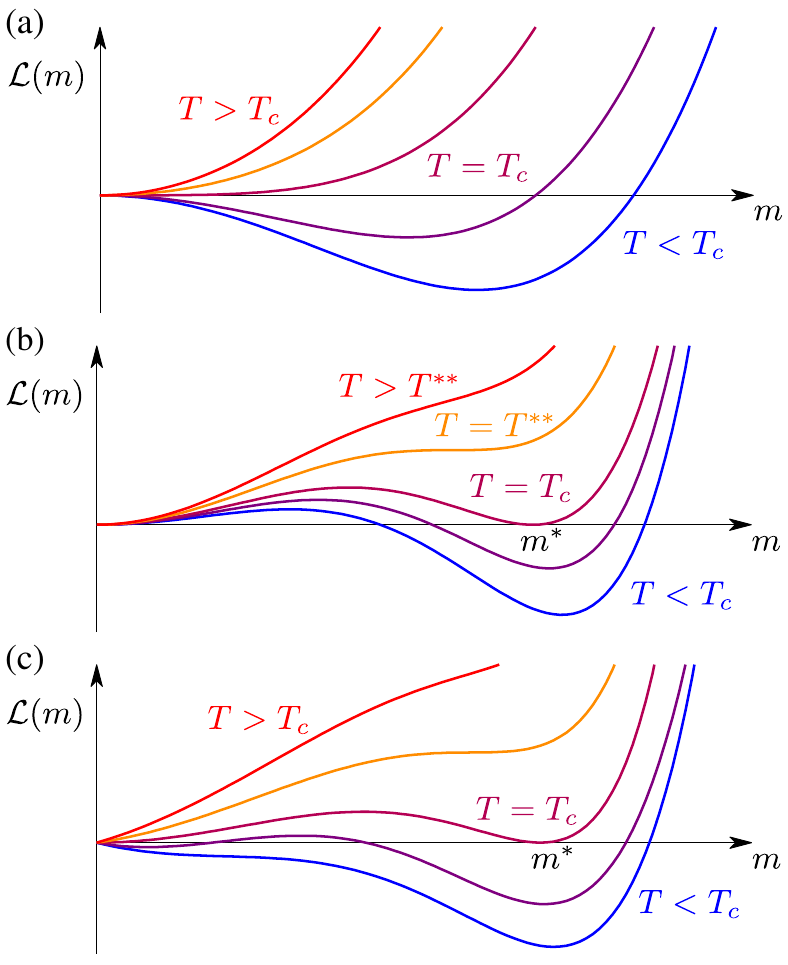}
\caption{Schematic plots of the Landau free energy $\mathcal{L}(m)$ as a function of the order parameter $m$ for (a) second-order, (b) first-order, and (c) hybrid phase transitions in thermal systems.
}
\label{fig:landau}
\end{figure}

Recently, the Landau theory was extended to the HPT. The authors of Ref.~\cite{at} investigated the AT model on scale-free networks. In the AT model, two types of Ising spins are located on each node of a scale-free network. Two spins of each type at the nearest-neighbor nodes interact with strength $J_2$, and four spins of both types at the nearest-neighbor nodes interact with strength $J_4$. The Landau free energy was established. Owing to the power-law behavior of the degree distribution of scale-free networks, the Landau free energy contains $m$ terms with non-integer powers. For specific cases in the parameter space ($T, J_4/J_2, \lambda)$, where $\lambda$ is the exponent of the degree distribution, an HPT occurs at the so-called critical endpoint. The order parameter jumps and includes critical behavior at the same transition point. The fluctuations of the order parameter are finite and diverge on either side of the transition point. The authors of Ref.~\cite{at} investigated the profile of the Landau free energy at this critical endpoint and established the criterion for the HPT within the Landau theoretical scheme as follows: At $T=T_c$, the free energy has two global minima at $m=0$ and $m^* > 0$. Thus, for $T > T_c$, the global minimum occurs at $m=0$, and for $T < T_c$, the global minimum occurs at $m^*(T)$. Mathematically, the criterion for the HPT is written as   
\begin{align} 
&\mathcal{L}=0,~\frac{\partial \mathcal{L}}{\partial m}=0,~{\rm and}~~\frac{\partial^2 \mathcal{L}}{\partial m^2}=0 &{\rm at}~m=0~~~&{\rm and}~T=T_c, \nonumber \\ 
&\mathcal{L}=0,~\frac{\partial \mathcal{L}}{\partial m}=0,~{\rm and}~~\frac{\partial^2 \mathcal{L}}{\partial m^2} \ge 0 &{\rm at}~m=m^*~&{\rm and}~T=T_c, \nonumber \\ 
&\mathcal{L} < 0,~\frac{\partial \mathcal{L}}{\partial m}=0,~{\rm and}~~\frac{\partial^2 \mathcal{L}}{\partial m^2} > 0 &{\rm at}~m=m^*~&{\rm and}~T < T_c.  
\label{criteria_hpt}
\end{align} 
The profiles of the Landau free energy as a function of the order parameter for different types of phase transitions are shown in Fig.~\ref{fig:landau}.
This criterion was confirmed by theoretical and experimental studies of the crystallization of colloidal magnets, in which the free energy is zero throughout the region $m=[0,m^*]$~\cite{colloid}. 

We note that the Landau theory criterion for the HPT was established in equilibrium thermal systems. Thus, in this paper, we aim to examine whether there exists a quantity corresponding to the Landau free energy and then to check whether the criterion for the HPT is still valid and useful in nonequilibrium dynamic systems. For this purpose, we consider Kuramoto models (KMs) with particular types of natural frequencies that exhibit HPTs~\cite{sync_pazo, sync_mendes, sync_basnarkov}. 
Contrary to infinite systems, for finite systems, there exist local minima at which the system is in metastable states. We investigate the stability of the metastable state as a function of system size and simulation time.

The paper is organized as follows: In Sec.~\ref{sec:kuramoto_model}, we describe the KM and its analytic approach to the self-consistency argument. In Sec.~\ref{sec:potential}, using an analogy with the effective free energy of the Landau theory, we introduce the {\sl ad hoc} potential from the self-consistency equation (SCE) and determine the types of synchronization transitions, including the HPT, for KMs with several types of natural frequencies in the thermodynamic limit. In Sec. \ref{sec:finite}, we extend the analysis to finite systems and compare the results with numerical simulations to verify the proposed scheme. We summarize and conclude with a discussion of our results in the final section.
 
\section{The Kuramoto model}\label{sec:kuramoto_model}
The KM describes a synchronization transition of $N$ oscillators in an all-to-all coupled system. Each phase evolves in time according to the Kuramoto equation (KE):
\begin{align}
\dot\theta_i = \omega_i+\frac{K}{N}\sum_{j=1}^N \sin (\theta_j-\theta_i),
\label{eq:kuramoto}
\end{align}
where $\theta_i$ denotes the phase of oscillator $i$, $\omega_i$ is its natural frequency drawn from a distribution $g(\omega)$, and $K$ is the coupling strength. 
The collective dynamics of the oscillators is quantified by the complex order parameter $Z$, which is defined as $Z=re^{i\psi}=\sum_{j=1}^N e^{i\theta_j}/N$, where $r$ is the phase coherence of the oscillators and serves as the order parameter during synchronization. $\psi$ is the average phase. 
Oscillators with natural frequencies satisfying $|\omega_{i}| \le Kr$ are phase-locked in the rotating frame with $\psi=0$. These oscillators contribute to the nonzero coherence $r$. In the steady state, the order parameter $r$ satisfies the SCE 
\begin{align}
r &= \int_{-{\pi}}^{{\pi}} d\theta \int_{-Kr}^{Kr} d\omega \cos\theta g(\omega) \delta \left(\theta - \arcsin \left( \frac{\omega}{Kr}\right) \right) \nonumber\\
&= \int_{-Kr}^{Kr} d\omega \sqrt{1-\frac{\omega^{2}}{K^{2}r^{2}}}g(\omega) \equiv f(r)\,.
\label{eq:sce}
\end{align}
Thus, the SCE is reduced to $F(r)\equiv f(r)-r=0$. 

\section{{\sl ad hoc} free energy}\label{sec:potential}

To investigate the dynamic flow and the stability of the SCE, one may choose an {\sl ad hoc} potential, which makes it possible to visualize the entire landscape in a given parameter space. Like the Landau theory, this landscape may give some clues to determining the synchronization transition types.

Let us define the {\sl ad hoc} potential $U(r)$ through the relation $F(r)=-dU(r)/dr$. In turn, $U(r)$ is written as 
\begin{align}
U(r) = \int_{0}^{r} (r^\prime-f(r^\prime))dr^\prime,
\label{eq:potential_sce}
\end{align}
where we set $U(0) = 0$ for simplicity. This suggests that for a given frequency distribution $g(\omega)$, $f(r)$ as defined in Eq. \eqref{eq:sce} allows us to explore the potential across the order parameter region. In the following, we investigate the profiles of {\sl ad hoc} potentials for second-order, first-order, and hybrid synchronization transitions for different types of natural frequency distributions.

\subsection{Second-order synchronization transition:\\ For the Gaussian distribution $g(\omega)$}\label{subsec:gaussian}

Here we consider the {\sl ad hoc} potential of the SCE for the Gaussian distribution $g(\omega)$ given by
\begin{align}
g(\omega) = \frac{1}{\sqrt{2\pi}}e^{-\frac{\omega^2}{2}}.
\label{eq:gaussian_dist}
\end{align}
We obtain the SCE as 
\begin{align}
r &= \sqrt{\frac{\pi A}{2}}e^{-A} \left[ I_{0} \left( A \right) + I_{1} \left( A \right) \right],
\label{eq:gaussian_sce}
\end{align}
where $A= K^2 r^2/4$, and $I_{\alpha}$ ($\alpha=0$ and $1$) denotes the modified Bessel functions of the first kind. For this $g(\omega)$, the order parameter increases continuously from $r=0$ to finite $r$ as $K$ is increased from a transition point $K_c=2/[\pi g(0)]$ \cite{kuramoto1, kuramoto2}. 

\begin{figure}
\includegraphics[width=.95\linewidth]{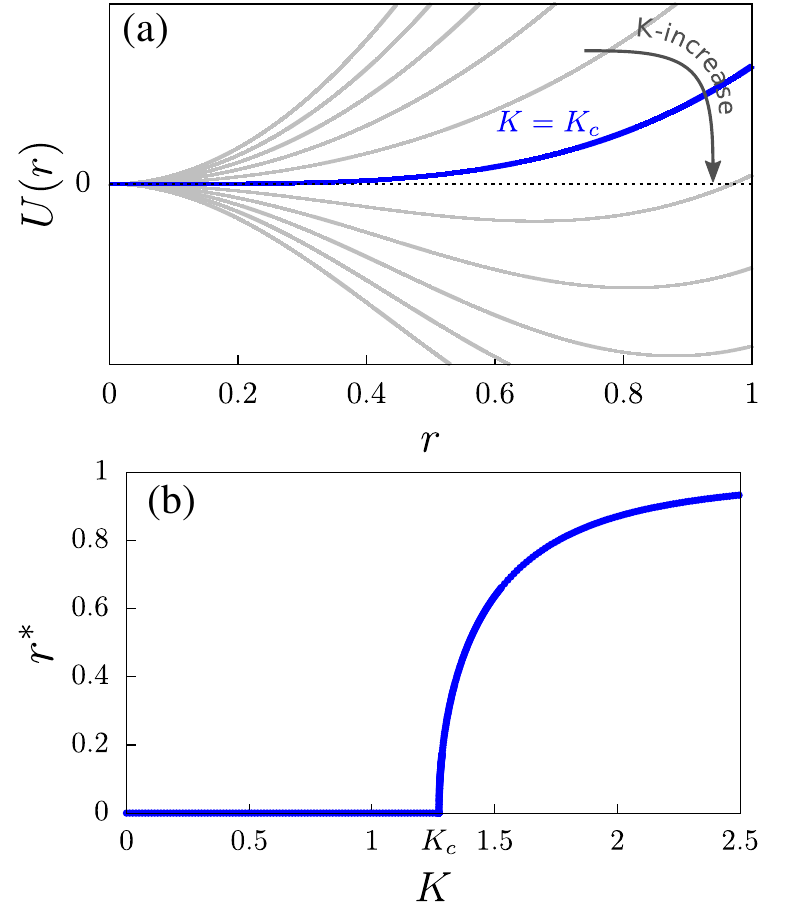}
\caption{
(a) Plot of ad hoc potential $U(r)$ given by Eq.~\eqref{eq:unimodal_potential} versus $r$. The potential exhibits a global minimum at $r=0$ for $K \le K_c$. For $K > K_c$, the position of the minimum increases continuously from 0 as $K$ is increased. (b) Plot of the position of the minimum of $U(r)$, denoted as $r^*$, versus $K$ for the Gaussian distribution $g(\omega)$. Here, a continuous transition occurs at $K_c$, and $r^*$ follows the formula \eqref{eq:unimodal_beta} above the critical point. 
}
\label{fig:gaussian_potential}
\end{figure}

Thus, we expand the r.h.s. of Eq.~\eqref{eq:gaussian_sce} with respect to $r$ at $r=0$ for $K=K_c$ and obtain that 
\begin{align}
r = \frac{K}{K_c}r-\frac{K^3}{8K_c}r^3+O(r^5) \,.
\label{eq:unimodal_sce}
\end{align}
The {\sl ad hoc} potential is obtained as 
\begin{align}
U(r) = \frac{K-K_c}{2K_c}r^2+\frac{K^3}{32K_c}r^4+O(r^6) \,.
\label{eq:unimodal_potential}
\end{align}
The profile of the {\sl ad hoc} potential is shown in Fig.~\ref{fig:gaussian_potential}(a) for various $K$ values. The sign of the coefficient of the $r^2$ term changes from positive to negative as $K$ is decreased beyond $K_c$, implying that the stability at $r=0$ is also inverted. We again investigate the relationship between the position of the minimum and the coupling strength, and obtain
\begin{align}
r \sim (K-K_c)^{1/2}
\label{eq:unimodal_beta}
\end{align}
for $K \rightarrow K_c$. Using numerics, we plot $r^*$, at which the minimum of $U(r)$ appears, in Fig. \ref{fig:gaussian_potential}(b) as a function of the coupling strength $K$. Starting from a small value of $K$, the minimum remains at $r=0$ until $K$ approaches $K_c$, and it increases continuously for $K>K_c$ following the relation given in Eq. \eqref{eq:unimodal_beta} \cite{kuramoto1, kuramoto2}.

\subsection{First-order synchronization transition} 

\subsubsection{When the degree and frequency are correlated on a scale-free network with  $2<\lambda<3$}\label{subsec:df_sf_es}

Refs.~\cite{sync_moreno,sync_mendes} consider the KM with degree--frequency correlation on scale-free networks with a power-law degree distribution $P_d(q)\sim q^{-\lambda}$. Using the annealed network approach, the KE is written as 
\begin{align}
\dot{\theta}_i = \omega_i+\sum_{j=1}^N \frac{K q_i q_j}{N\langle q \rangle}\sin(\theta_j-\theta_i),
\label{eq:kuramoto_sf2}
\end{align}
where $q_i$ and $q_j$ are the degrees of nodes $i$ and $j$, respectively, and $\langle q \rangle$ is the mean degree, which is defined as $\langle q \rangle = \sum_j q_j/N$. The degree--frequency correlation is given in the form of $\omega_i=q_i$. 
The complex order parameter of the system is defined as
\begin{align}
Z = re^{i\psi}=\frac{1}{N\langle q \rangle}\sum_{l=1}^N q_l e^{i\theta_l},
\label{eq:orderparameter_sf1}
\end{align}
where $r$ is the coherence, and $\psi$ is the average phase. 

It was shown that when the degree exponent $\lambda$ is in the range $2< \lambda < 3$, the synchronization transition is first-order~\cite{sync_mendes}. Following the steps taken in Ref.~\cite{sync_mendes}, one can obtain the SCEs for two parameters, $\alpha \equiv rK$ and the group angular velocity, $\Omega$, as 
\begin{align}
\langle q \rangle - \Omega &= \int_{1}^{\infty} dq P_d(q) (q-\Omega) \sqrt{1-\left(\frac{\alpha q}{q-\Omega} \right)^2} \times \nonumber\\
&\Theta \left(\left| \frac{q-\Omega}{\alpha q} \right|-1 \right)
\label{eq:df_imaginary}
\end{align}
and
\begin{align}
r &= \frac{\alpha}{K} \nonumber\\
&= \frac{1}{\langle q \rangle} \int_{1}^{\infty} dq P_d(q) q \sqrt{1-\left(\frac{q-\Omega}{\alpha q} \right)^2} \Theta \left(1-\left| \frac{q-\Omega}{\alpha q} \right| \right) \,.
\label{eq:df_real}
\end{align}

\begin{figure}
\includegraphics[width=.95\linewidth]{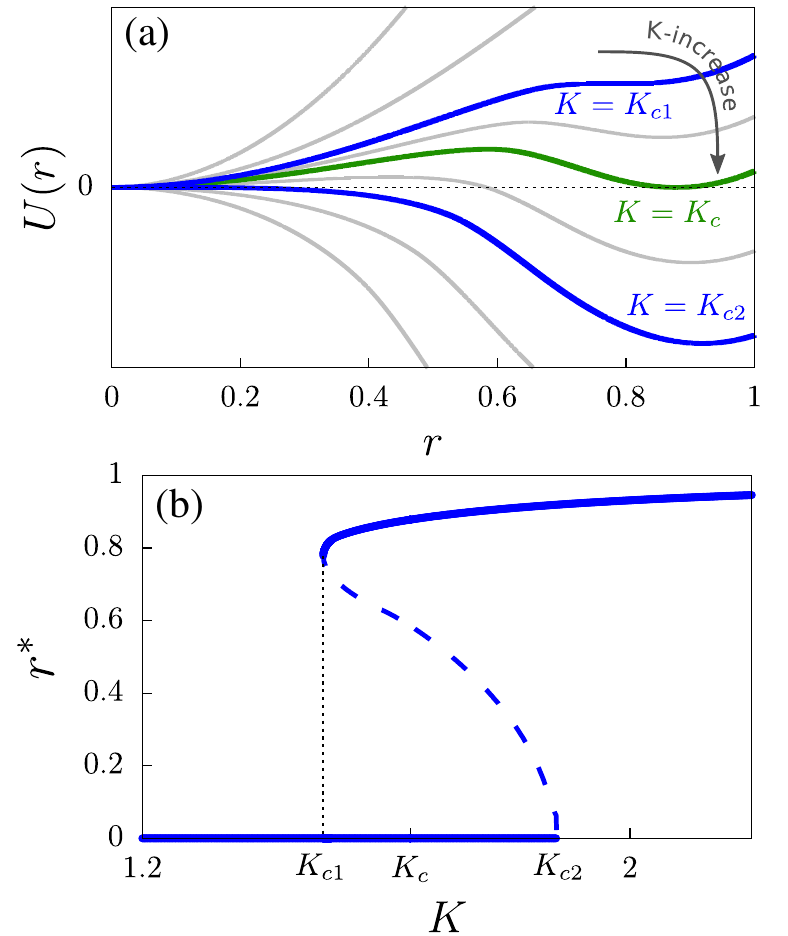}
\caption{
(a) {\sl Ad hoc} potential shape $U(r)$ for different $K$ values for the KE given by Eq.~\eqref{eq:kuramoto_sf2} on scale-free networks with degree exponent $\lambda=2.8$. The potential exhibits only a minimum at $r^*=0$ for $K<K_{c1}$. As $K$ is increased, another local minimum is generated at $r^* >0$ for $K>K_{c1}$. As $K$ is further increased, $U(r)$ becomes smaller at this local minimum position; eventually, when $K=K_c$, $U(r)$ becomes zero at a certain $r^*>0$. Thus, there exist two global minima at $r^*=0$ and $r^* >0$. The minimum at $r^* > 0$ becomes only a global minimum as $K$ is further increased. By contrast, the minimum at $r=0$ disappears when $K=K_{c2}$. (b) Position $r^*$ at which $U(r)$ becomes either a local or a global minimum in (a) as a function of $K$. $r^*$ exhibits a discontinuous transition in the region [$K_{c1}$, $K_{c2}$]. Blue dashed curve is the trajectory of the local maximum position of $U(r)$ as $K$ is increased indicating an unstable curve.
}
\label{fig:es_df}
\end{figure}

For $2<\lambda<3$, by solving SCEs \eqref{eq:df_imaginary} and \eqref{eq:df_real} for $\alpha$ and $\Omega$, one can evaluate the {\sl ad hoc} potential. Because it is not as simple to calculate analytically, we first obtained the solution of $\Omega(\alpha)$ from Eq.~\eqref{eq:df_imaginary} numerically and then solved for the SCE by substituting it into Eq.~\eqref{eq:df_real}. As shown in Fig.~\ref{fig:es_df}(a), in this case, $U(r)$ exhibits a minimum at $r=0$ when $K<K_{c1}$ and two minima at $r=0$ and $r>0$ when $K_{c1}<K<K_{c2}$, where $K_{c1}$ and $K_{c2}$ are defined in the caption of Fig.~\ref{fig:es_df}. As $K$ is increased beyond $K_c$ defined in the caption of Fig.~\ref{fig:es_df}, the minimum at $r>0$ becomes a global minimum. As $K$ is further increased to $K > K_{c2}$, the minimum at $r=0$ no longer exists. This change in the potential shape as a function of $K$ provides an intuitive understanding of the first-order synchronization transition as it appears for the first-order transition in the Landau theory for thermal systems. The order parameter behaves as shown in Fig. \ref{fig:es_df}(b).

\subsubsection{When the interaction strength depends on the frequency}\label{subsec:es_gn}
Another model exhibiting a first-order synchronization transition, the explosive synchronization model, was introduced in Ref.~\cite{sync_es_gn}. The model equation is written as
\begin{align}
\dot{\theta}_j &= \omega_j+ \frac{K |\omega_j|}{\sum_{l=1}^N A_{jl}} \sum_{l=1}^N A_{jl}\sin (\theta_l-\theta_j),
\label{eq:kuramoto_gn1}
\end{align}
where $A_{jl}$ denotes an element of the adjacency matrix. The complex order parameter is defined as
\begin{align}
Z = re^{i\psi}=\frac{1}{N}\sum_{l=1}^N e^{i\theta_l} \,.
\label{eq:orderparameter_gn1}
\end{align}
\begin{figure}
\includegraphics[width=.95\linewidth]{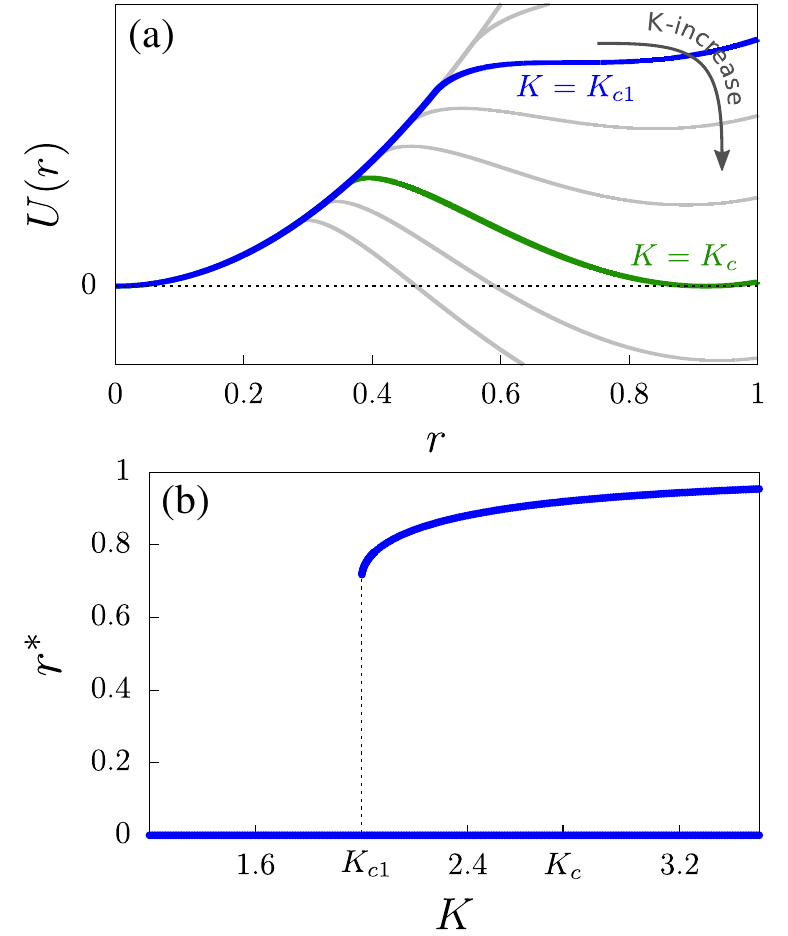}
\caption{(a) {\sl Ad hoc} potential $U(r)$ given in Eq.~\eqref{eq:Ur_symm}. The potential exhibits one minimum at $r=0$ for $K<K_{c1}$. As $K$ is increased, local minimum develops at $r^* >0$ for $K>K_{c1}$, but it is not a global minimum yet. As $K$ is increased further, $U(r)$ at the minimum point becomes smaller. When $K=K_c$, $U(r)$ becomes zero at both $r=0$ and $r^*>0$. So $U(r)$ at $r^* >0$ reaches a global minimum as $K$ is further increased. Note that minimum of $U(r)$ at $r=0$ remains as long as $K$ is finite. (b) Values of $r^*$ versus $K$.}
\label{fig:es_gn_symm}
\end{figure}
$\{\omega_i\}$ has a distribution $g(\omega)$. For symmetric $g(\omega)$ in all-to-all networks, one can obtain the equation
\begin{align}
\Delta \dot{\theta}_j &= \omega_j - K |\omega_j| r \sin(\Delta \theta_j),
\label{eq:reducedkuramoto_symm}
\end{align}
where $\Delta \theta_j \equiv \theta_j - \psi$, by following the derivation in Ref. \cite{sync_es_gn}.
When all the oscillators are phase-locked, i.e., $\Delta \dot{\theta}_j=0$ for all $j$, the solution is obtained as 
\begin{align}
\Delta \theta_j 
&= \begin{cases} 
\arcsin \left( \frac{1}{Kr} \right) &\textrm{for} \quad \omega_j > 0 \\
\arcsin \left( -\frac{1}{Kr} \right) &\textrm{for} \quad \omega_j < 0 \,.
\end{cases}
\label{eq:phase_symm}
\end{align}
From Eq. \eqref{eq:orderparameter_gn1}, the SCE can be written as
\begin{align}
r = &\frac{1}{2}\int_{-\pi}^{\pi}d\theta \int_{-\infty}^{\infty} d\omega g(\omega) \cos\theta ~\Theta \left(\omega-\left| \frac{\omega}{Kr} \right| \right) \times \nonumber \\ &\left( \delta \left( \theta-\arcsin \left( \frac{1}{Kr} \right) \right) + \delta \left( \theta-\arcsin \left( -\frac{1}{Kr} \right) \right) \right) \\
= &\sqrt{1- \left( \frac{1}{Kr} \right)^2} \Theta \left(1-\left| \frac{1}{Kr} \right| \right) \,.
\label{eq:sce_symm}
\end{align}
Hence, $f(r)$ is determined as follows: 
\begin{align}
f(r) &= \begin{cases} 
0 &\textrm{for} \quad Kr \le 1 \\
\sqrt{1- \left( \frac{1}{Kr} \right)^2} &\textrm{for} \quad Kr > 1 \,.
\end{cases}
\label{eq:fr_symm}
\end{align}
Accordingly, $U(r)$ is obtained as follows:
\begin{align}
U(r) &= \begin{cases} 
\frac{r^2}{2} &\textrm{for} \quad r \le \frac{1}{K} \\
\frac{r^2}{2} - r \sqrt{1- \left( \frac{1}{Kr} \right)^2} \\ - \frac{2}{K} \arctan \left( \sqrt{\frac{Kr+1}{Kr-1}} \right) &\textrm{for} \quad r > \frac{1}{K} \,.
\end{cases}	
\label{eq:Ur_symm}
\end{align}

Fig. \ref{fig:es_gn_symm} shows a discontinuous transition at $K=K_{c1}$, from which a minimum of $U(r)$ at $r>0$ starts to develop. At $K_c$, the minimum of $U(r)$ becomes zero for $r^* > 0$. As $K$ is increased further, this minimum at $r^* > 0$ is a global minimum. Note that unlike the case in the previous subsection, $K_{c2}$, at which the second derivative of $U(r)$ with respect to $r$ at $r=0$ becomes zero, does not exist. It is always positive as long as $K$ is finite. Therefore, there is no hysteresis curve. In the limit $K \rightarrow \infty$, the minimum at $r^*>0$ becomes the dominant solution, and the minimum at $r=0$ disappears.

\subsection{Hybrid synchronization transition}\label{subsec:hybrid}

\subsubsection{For a uniform distribution $g(\omega)$}\label{subsec:uniform}
We consider the {\sl ad hoc} potential for the uniform distribution $g(\omega)$ given by 
\begin{align}
g(\omega) = \begin{cases} \frac{1}{2\gamma} \quad &\textrm{for} \quad |\omega| \le \gamma\\
0 \quad &\textrm{for} \quad |\omega| > \gamma
\end{cases},
\label{eq:uniform_dist}
\end{align}
where $\gamma$ is the half-width of the distribution. Thus, $f(r)$ becomes
\begin{align}
f(r) = \begin{cases} \frac{K}{K_c}r &\textrm{for} \quad Kr \le \gamma \\
        \frac{1}{2}\sqrt{1-\frac{\gamma^{2}}{K^{2}r^{2}}}+\frac{Kr}{2\gamma}\arcsin\left(\frac{\gamma}{Kr}\right) &\textrm{for} \quad Kr > \gamma
\end{cases},
\label{eq:fr_cases}
\end{align}
where $K_c$ is the transition point in the thermodynamic limit, determined by $K_c r_c = \gamma$~\cite{sync_pazo}. 
Explicitly, $K_c =\frac{4\gamma}{\pi}$, and $r_c =\frac{\pi}{4}$.
The potential is determined as 
\begin{align}
U(r) = \begin{cases} 
\frac{K_c - K}{2K_c}r^2 &\textrm{for} \quad r \le \frac{\gamma}{K} \\
\int_{0}^{r} ( r^\prime-\frac{1}{2}\sqrt{1-\frac{\gamma^{2}}{K^{2}r^{\prime 2}}} \\
\quad -\frac{Kr^\prime}{2\gamma}\arcsin( \frac{\gamma}{Kr^{\prime}}) ) dr^\prime &\textrm{for} \quad r > \frac{\gamma}{K}
\end{cases}.
\label{eq:potential_uniform}
\end{align}

\begin{figure}
\includegraphics[width=.95\linewidth]{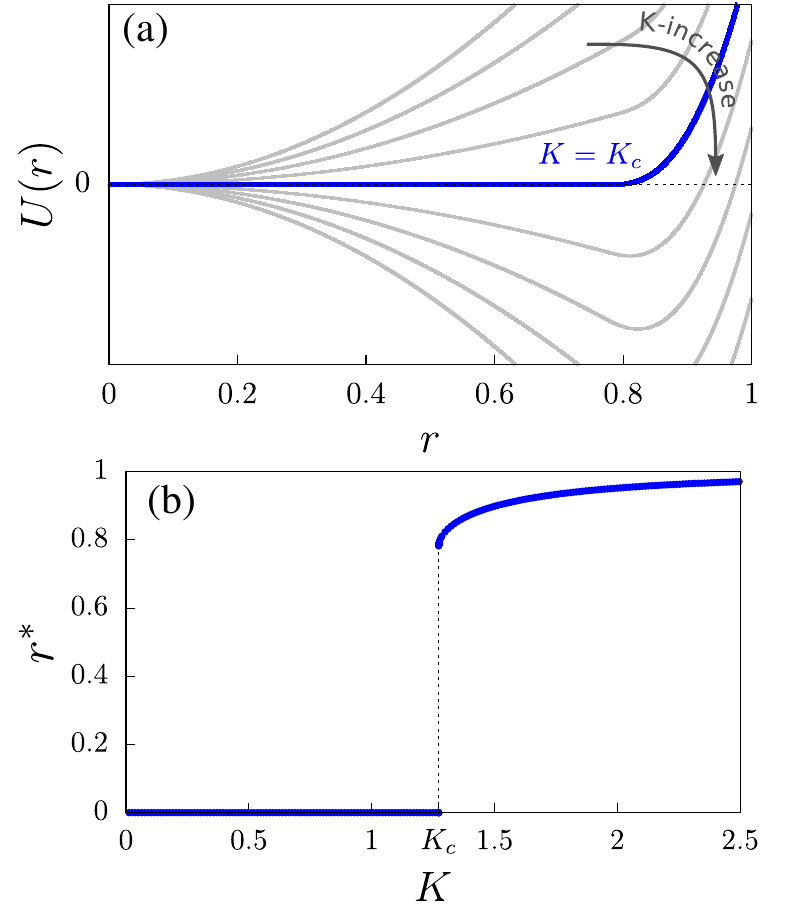}
\caption{
(a) Plot of {\sl ad hoc} potential $U(r)$ as a function of the order parameter $r$ for the case \eqref{eq:potential_uniform}. The potential exhibits a global minimum at $r=0$ for $K < K_c$ and forms a plateau at $K=K_c$ in the region $r \in [0, r_c=\pi/4]$. When $K>K_c$, a global minimum appears above $r_c=\pi/4$, and it increases gradually as $K$ is increased further. (b) Plot of $r^*$ values, at which global minima of $U(r)$ are positioned for given $K$s, as a function of $K$ for the uniform distribution $g(\omega)$. $r^*$ undergoes a discontinuous transition at $K=K_c$ and follows Eq.~\eqref{eq:uniform_beta} above the critical point $K_c$.
}
\label{fig:uniform_potential}
\end{figure}

Numerical evaluations of $U(r)$ for arbitrary values of $K$ are plotted in Fig.~\ref{fig:uniform_potential}(a). For $K<K_c$, the coefficient of $r^2$ is positive for $r \le \gamma/K$ in Eq.~\eqref{eq:potential_uniform}, so the solution at $r=0$ is stable. As $K$ is increased, the coefficient approaches zero, and the potential becomes flatter near the origin. At $K=K_c$, the coefficient becomes zero, and a plateau is formed across the range $r \le \gamma/K_c = r_c$, as shown in Fig. \ref{fig:uniform_potential}(a). When $K > K_c$, the coefficient is negative, and thus the solution $r=0$ becomes unstable. In this case, a stable minimum emerges in the region $r > r_c$.

The minimum of the potential in the region $r > r_c$ can be calculated by performing an expansion above both $K_c$ and $r_c$ as $K=K_c+\epsilon$ and $r=r_c+\delta$. 
By substituting these expressions into Eq.~\eqref{eq:potential_uniform} and taking the limit $\epsilon$ and $\delta \rightarrow 0$, we obtain the potential for $r > \gamma/K$,
\begin{align}
U(r) = -\frac{\pi^2\epsilon}{16\gamma}\delta+\frac{32\sqrt{2}}{15\pi^{3/2}}\delta^{5/2}+{\rm h.o.} \,
\label{eq:uniform_expand}
\end{align}
Minimizing the potential ($dU(r)/dr = 0$), we obtain the hybrid synchronization transition behavior of the order parameter as  
\begin{align}
r-r_c = \left(\frac{9\pi^7}{2^{17}\gamma^2}\right)^{1/3}(K-K_c)^{2/3} \,.
\label{eq:uniform_beta}
\end{align}
The stable fixed point of the order parameter follows this relation, which is consistent with the result obtained in \cite{sync_pazo}. 
Fig.~\ref{fig:uniform_potential}(b) shows that the position of the minimum $r^*$ exhibits a discontinuous jump at the critical value $K_c$. 
We remark that the potential $U(r)$ satisfies the Landau criterion for the HPT in thermal systems given in Eq.~\eqref{criteria_hpt}.

\subsubsection{When the degree and frequency are correlated on scale-free networks with $\lambda=3$}\label{subsec:df_sf}
Here we consider the Kuramoto dynamics on scale-free networks that exhibit a power-law degree distribution $P_d(q)\sim q^{-\lambda}$, where $q$ denotes the degree, for $\lambda=3$. In this case, the KE is known to exhibit a hybrid synchronization transition~\cite{sync_moreno,sync_mendes}. The KE is written as Eq.~\eqref{eq:kuramoto_sf2}.
In particular, the condition $\omega_i=q_i$, at which a hybrid synchronization transition occurs, is given. 

The SCE for a scale-free network with $\lambda=3$ was derived in the appendix of Ref. \cite{sync_mendes}:
\begin{align}
r = \frac{1}{2} \int_{-1}^{1} dx \sqrt{1-\left(\frac{x}{\alpha} \right)^2} \Theta \left(1-\left| \frac{x}{\alpha} \right| \right) \,.
\label{eq:sce2}
\end{align}
By using $\alpha=rK$, the equation can be written as
\begin{align}
r &= \begin{cases} 
\frac{K}{K_c}r &\textrm{for} \quad Kr \le 1 \\
\frac{1}{2} \sqrt{1-\frac{1}{K^2r^2}} + \frac{Kr}{2}\arcsin \left(\frac{1}{Kr}\right) &\textrm{for} \quad Kr > 1
\end{cases},
\label{eq:sce3}
\end{align}
where $K_c=4/ \pi$, and $r_c=\pi/4$. This result is reduced to the same as that for the uniform distribution of $g(\omega)$ in all-to-all connected networks discussed in Sec. \ref{subsec:uniform}. Therefore, one can obtain exactly the same potential $U(r)$ as that given for the uniform frequency distribution $\gamma=1$.

\subsubsection{For a flat distribution with exponential tails}\label{subsec:basnarkov} 

\begin{figure}[h!]
\includegraphics[width=.95\linewidth]{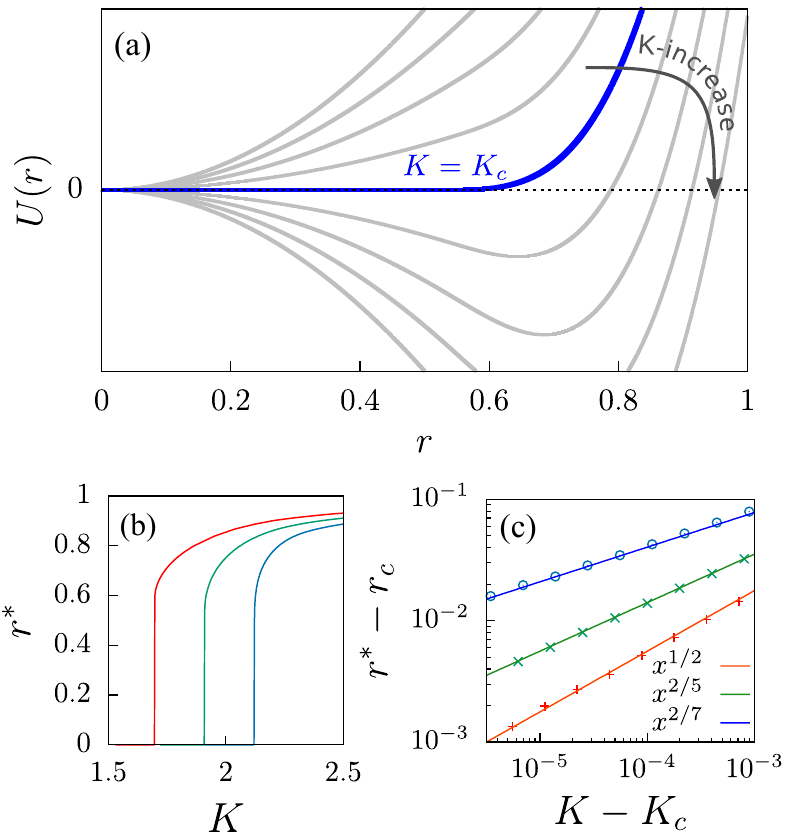}
\caption{
(a) Plot of {\sl ad hoc} potential $U(r)$ as a function of $r$ for the case \eqref{eq:basnarkov_fr} with $m=1$. The potential exhibits a plateau in the region $[0, r_c]$. (b) Plot of $r^*$ values at which minima of $U(r)$ are positioned as a function of $K$ for various sets of $(m, \alpha)$. From the left, the curves are for ($m,\alpha$) = (0.5,1) (in red), (1,1) (in green), and (2,1) (in blue). For all cases, $r^*$ undergoes a discontinuous transition at the critical point $K_c$. (c) Above $K_c$, the exponent $\beta$ of the order parameter is measured for $m=0.5$ (red), 1 (green), and 2 (blue). The straight lines are drawn according to the theoretical formula [Eq.~\eqref{eq:basnarkov_beta}] for each case.
}
\label{fig:basnarkov_fig}
\end{figure}

In Ref.~\cite{sync_basnarkov}, the uniform natural frequency distribution was extended by adding tails on each side as follows: 
\begin{align}
g(\omega) &=\begin{cases} g(0) &\textrm{for} \quad |\omega|\leq \alpha\\ g(0)[1-c(|\omega|-\alpha)^m] &\textrm{for} \quad \alpha \le |\omega| \le \alpha+c^{-\frac{1}{m}} \\ 
 0 &\textrm{otherwise}
\end{cases},
\label{eq:basnarkov_dist}
\end{align}
where $c$ is a positive constant, and $g(0)$ is given by
\begin{align}
g(0)=\frac{1}{2} \frac{1}{\alpha+[m/(m+1)]c^{-1/m}}
\label{eq:basnarkov_norm}
\end{align}
according to the normalization condition. For this distribution, we obtain $f(r)$ as
\begin{widetext}
\begin{align}
f(r) &=\begin{cases} \frac{K}{K_c}r &\textrm{for} \quad Kr\leq \alpha\\ \frac{K}{K_c}r-2g(0)c \int_{\alpha}^{Kr} \left( \sqrt{1-\frac{\omega^2}{K^2r^2}}(\omega-\alpha)^m \right) d\omega &\textrm{for} \quad \alpha < Kr \le \alpha+c^{-\frac{1}{m}} \\
\frac{K}{K_c}r-2g(0)c \int_{\alpha}^{\alpha+c^{-\frac{1}{m}}} \left( \sqrt{1-\frac{\omega^2}{K^2r^2}}(\omega-\alpha)^m \right) d\omega &\textrm{for} \quad \alpha+c^{-\frac{1}{m}} < Kr \end{cases},
\label{eq:basnarkov_fr}
\end{align}
\end{widetext}
where $K_c=2/[\pi g(0)]$. $U(r)$ was also calculated numerically using Eq.~\eqref{eq:basnarkov_fr}, as shown in Fig. \ref{fig:basnarkov_fig}(a). As in previous sections, plateau region of $g(\omega)$ leads the system to exhibit a hybrid synchronization transition with a flat potential at the critical point. A calculation of $r^*$ for the potential confirms that
\begin{align}
r-r_c \sim (K-K_c)^{2/(2m+3)},
\label{eq:basnarkov_beta}
\end{align}
which was studied in Ref.~\cite{sync_basnarkov}. When $m=0$, the exponent $\beta$ becomes $2/3$, which is consistent with that of the uniform distribution. 

\subsubsection{For a flat distribution with power-law tails}\label{subsec:trim_tail}

\begin{figure*}[t]
	\includegraphics[width=\textwidth]{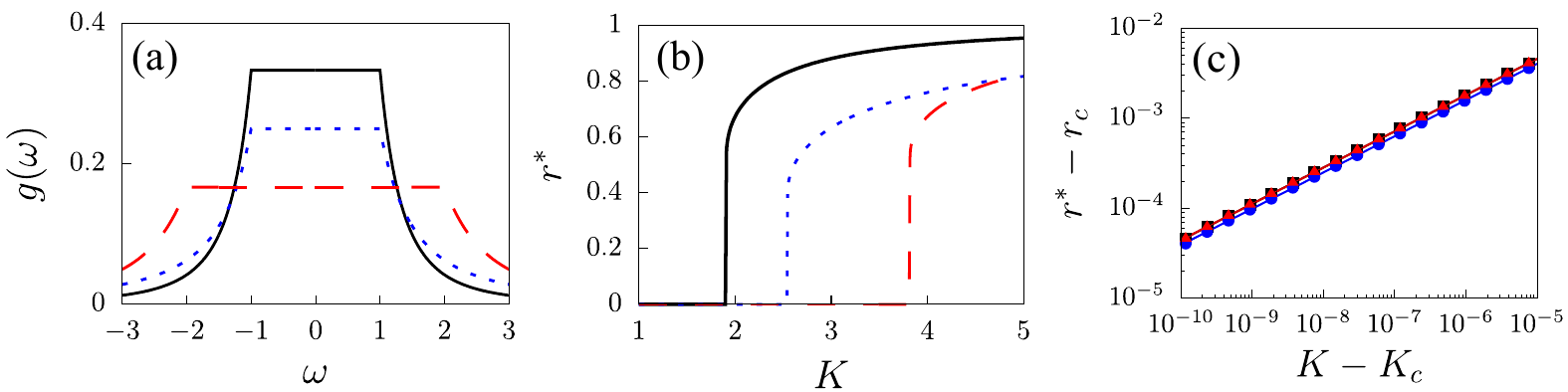}
	\caption{ (a) Flat-with-tails distribution for various values of $(\alpha,m)$: solid, $(1,3)$; dotted, $(1,2)$; and dashed, $(2,3)$. (b) Order parameter curve obtained using the SCE. (c) $\beta = 0.40$ is measured for all cases.}
\end{figure*}

We consider the Lorentzian distribution with an upper cutoff defined as 
\begin{align}
g(\omega) &=\begin{cases} g(0) & |\omega|\leq \alpha\\ \frac{1}{\mathcal{N}}\frac{\gamma/\pi}{\gamma^2+\omega^2} & |\omega|>\alpha\end{cases},
\end{align}
where the normalization is calculated as $\mathcal{N} = 1 - (2/\pi)\arctan(\alpha/\gamma) + 2\gamma\alpha/(\pi(\gamma^2+\alpha^2))$, and 
\begin{align}
g(0)=\frac{1}{\mathcal{N}}\frac{\gamma /\pi}{\gamma^2+\alpha^2}.
\end{align}
$g(\omega)$ is thus flat in $(-\alpha,\alpha)$ and has a long-decay tail $\sim |\omega|^{-2}$ on each side. We find a universal hybrid critical exponent $\beta=2/5$, together with a plateau of $U(r)$ similar to that in Fig.~\ref{fig:basnarkov_fig}(a), for this distribution and for any flat distribution with power-law tails. 


Now, we consider a $g(\omega)$ that is flat in the interval $[-\alpha,\alpha]$ and decays in a power-law manner, $\sim |\omega|^{-m}$ ($m>1$), for $\omega > \alpha$. 
\begin{align}
g(\omega) = \begin{cases} g(0), & |\omega|\leq \alpha\\ \frac{g(0)\alpha^m}{|\omega|^m}, & |\omega|>\alpha \end{cases},
\end{align}
where
\begin{align}
    g(0)=\frac{m-1}{2\alpha m}
\end{align}
by the normalization condition, $\int g(\omega) d\omega =1$. For $Kr\geq\alpha$, the SCE is written as
\begin{widetext}
\begin{align}
    r &= Krg(0)\int_{-\alpha/(Kr)}^{\alpha/(Kr)} dx \sqrt{1-x^2} +\frac{2g(0)\alpha^m}{(Kr)^{m-1}}\int_{\alpha/(Kr)}^{1} x^{-m}\sqrt{1-x^2} dx \\
    &= 2Kr g(0) \int_0^{\theta_0} d\theta \cos^2\theta + \frac{2g(0)\alpha^m}{(Kr)^{m-1}}\int_{\sin^2\theta_0}^{1} \frac{1}{2} y^{-\frac{m+1}{2}}(1-y)^{\frac{1}{2}}dy. \label{eq:38}
\end{align}
\end{widetext}
where $\alpha/Kr \equiv \sin\theta_0$. Notice that at $\theta_0=\pi/2$, the SCE shows that the order parameter jumps by as much as $r_c={\pi g(0)\alpha}/{2}$ at $K_c={2}/{(\pi g(0))}$.  

Using the SCE, we obtain that 
\begin{align}
r-r_c \sim \frac{\alpha}{2K_c} \left(\frac{15\pi}{4mK_c}\right)^{2/5}(K-K_c)^{2/5}
\end{align}
within the leading order. Therefore, the transition is hybrid, and the associated exponent is $\beta=2/5$, which differs from the value of $\beta=2/3$ for the uniform distribution. 
Notice that in the limit $m\rightarrow\infty$, the second term of Eq.~\eqref{eq:38} vanishes because $\sin \theta_0 =\sin\left(\frac{\pi}{2}-\delta\theta\right)<1$, and thus $\beta=2/3$ is recovered.

\section{finite systems}\label{sec:finite}
In finite systems, the SCE given in Eq.~\eqref{eq:sce} is written as
\begin{align}
r = \frac{1}{N}\sum_{|\omega_i| \le Kr} \sqrt{1-\frac{\omega_{i}^{2}}{K^{2}r^{2}}}\equiv f(r) \,.
\label{eq:sce_finite}
\end{align}
This SCE may also be written in the form ${x}/{K}=h(x)$, where 
\begin{align}
h(x) \equiv \frac{1}{N}\sum_{|\omega_i| \le x} \sqrt{1-\frac{\omega_{i}^{2}}{x^2}}=\frac{x}{K} \,,
\label{eq:hx_define}
\end{align}
where $x \equiv Kr$ and $h(x)$ replaces $f(r)$. Here we consider that $g(\omega)$ is uniform. The {\sl ad hoc} potential is defined as it was above: 
\begin{align}
U(r) = \int_{0}^{r} (r^\prime-f(r^\prime))dr^\prime. \nonumber
\end{align}
We consider two cases in which the natural frequencies of each oscillator are taken randomly and regularly. 

\subsection{Random sampling of $\{\omega_i\}$}\label{subsec:random}
\begin{figure}
\includegraphics[width=.95\linewidth]{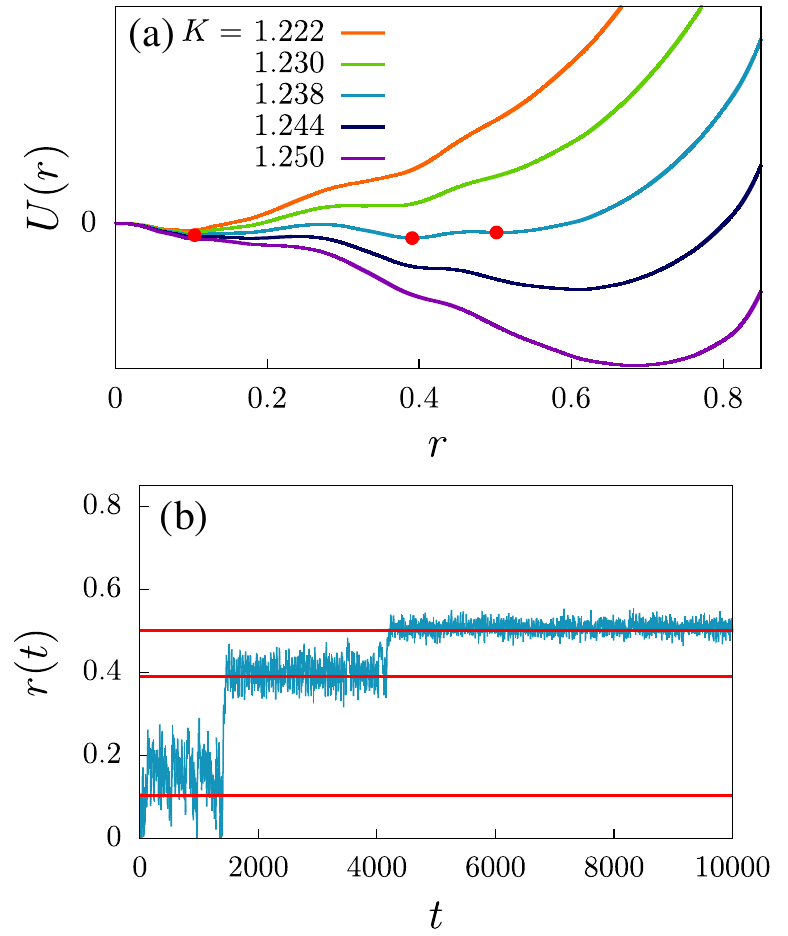}
\caption{(a) Plot of {\sl ad hoc} potential $U(r)$ as a function of $r$ for the case of random sampling of $\{\omega_i\}$ from the uniform distribution $g(\omega)$ with $\gamma=1$ for $K = 1.222, 1.230, 1.238, 1.244$, and $1.250$. Except the minimum at $r=0$, local minima of $U(r)$ are marked by red dots for $K =1.238$. (b) Time evolution of the order parameter $m(t)$ for the set of oscillators used in (a) with $K = 1.238$ and random initial phases. The three local minima positions in (a) are indicated by red lines.
}
\label{fig:random}
\end{figure}

We first consider the case that $\omega_i$ is selected randomly from the uniform distribution $g(\omega)$ given by~\eqref{eq:uniform_dist} for half of the oscillators ($i=1, \ldots, N/2$), and the other half are assigned values following $\omega_i = -\omega_{N-i+1}$ for $i=N/2+1, \ldots, N$, so that the mean natural frequency becomes zero. The {\sl ad hoc} potential of the SCE for each case is obtained as shown in Fig.~\ref{fig:random}(a). For a given $K$, there exist local minima, which are stable solutions of the SCE. The global minimum of the potential develops from $r=0$ as $K$ is increased, leading the order parameter to jump to a finite value. This abrupt change of the position of the global minimum suggests the possibility of a hybrid synchronization transition in the limit $N \rightarrow \infty$, as discussed in Sec.~\ref{subsec:uniform}.

To validate this scheme in view of the effective potential, we perform simulations for a system size  $N=6400$ using the fourth-order Runge--Kutta method up to $t=10^6$ time in steps of $\delta t=10^{-2}$. 

Fig.~\ref{fig:random}(b) shows the evolution of the order parameter under the same condition used in Fig.~\ref{fig:random}(a). Because initial phases of each oscillators are distributed randomly, the order parameter is $r \sim O(N^{-1/2})$ at $t=0$. As time runs, the dynamics proceeds according to the effective potential landscape. As shown in Fig.~\ref{fig:random}(b), $r(t)$ exhibits a plateau with some fluctuations for a certain period of time. Comparing with the profile of the {\sl ad hoc} potential, this pattern results from that the system is confined in a corresponding potential well before jumping to the next.

As $r$ increases, the number of drifting oscillators decreases according to $\langle N_d\rangle= (1-r/r_c)N$ and so do the dynamic fluctuations of the order parameter [see the width of the fluctuations in Fig. \ref{fig:random}(b)]. Moreover, the potential barrier from $r\approx 0.4$ to the left is higher than that to the right in Fig. \ref{fig:random}(a)], and thus the system tends to move to the right side of the landscape (larger $r$). Consequently, the system beginning at $r \sim O(N^{-1/2})$ passes through metastable states of local potential wells and then reaches the steady state, which corresponds to the rightmost position, as far as possible, among the positions of the local minima.

\begin{figure}
\includegraphics[width=.95\linewidth]{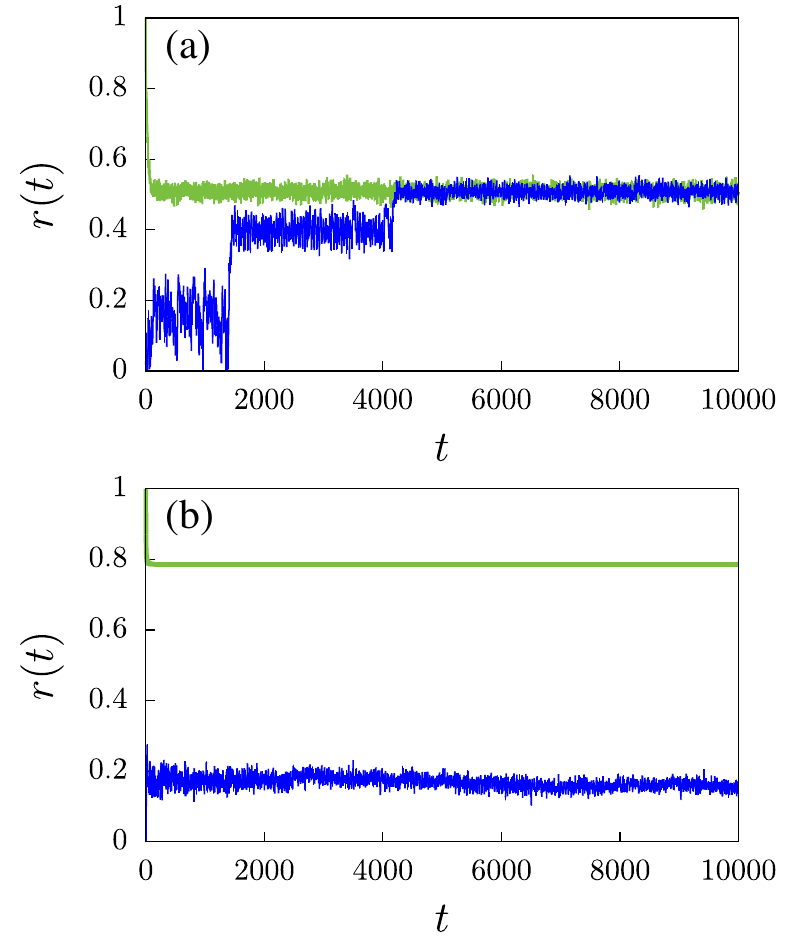}
\caption{Comparison of the order parameter behavior as a function of $t$ from different initial configurations with $r(0) \sim O(N^{-1/2})$ (blue, dark) and 1 (green, gray). Natural frequencies of each oscillator are selected randomly in (a) and regularly in (b). Numerical simulations are performed for the case $N=6400$ at $K = 1.238$ (a) and $K_c$ (b).
}
\label{fig:compare_dynamics}
\end{figure}

We also consider the evolution of the order paramter from different initial values of $r$. For the same  randomly sampled set used in Fig. \ref{fig:random}, the dynamics begins in a totally synchronized state, $r=1$, and flows to the steady state coinciding with the final state of the dynamics beginning at $r \sim O(N^{-1/2})$. This can be intuitively understood in terms of the {\sl ad hoc} potential shown in Fig. \ref{fig:random}(a). At the initial value of $r=1$, the dynamics of the system moves down from the far right side of the potential and first encounters a minimum at a certain value of $r$. As the fluctuations at this minimum are not sufficiently large to overcome the barrier on the left, the dynamics remains at this minimum, as shown in Fig.~\ref{fig:compare_dynamics}(a). This result does not differ much from that of other general random sets of $\{\omega_i\}$. 

\subsection{Regular sampling of $\{\omega_i\}$}\label{subsec:regular}

We consider that $\{\omega_i\}$ is selected regularly from the uniform distribution given in Eq.~\eqref{eq:uniform_dist}. In this case, $\omega_i$ is given as 
\begin{align}
\omega_i = -\gamma + \frac{\gamma}{N}(2i-1)
\label{eq:midpoint_rule}
\end{align}
for $i=1, \ldots, N$.
The SCE of Eq.~\eqref{eq:hx_define} is rewritten as 
\begin{align}
\frac{x}{K} = \frac{1}{N}\sum_{i=N-n+1}^n \sqrt{1-\frac{\omega_{i}^{2}}{x^2}}=h(x),
\label{eq:regular_finite_sce}
\end{align}
where $n$ is the index satisfying $\omega_n \le x < \omega_{n+1}$, so phase-locked oscillators contribute to the summation. 

\begin{figure}[h!]
\includegraphics[width=.95\linewidth]{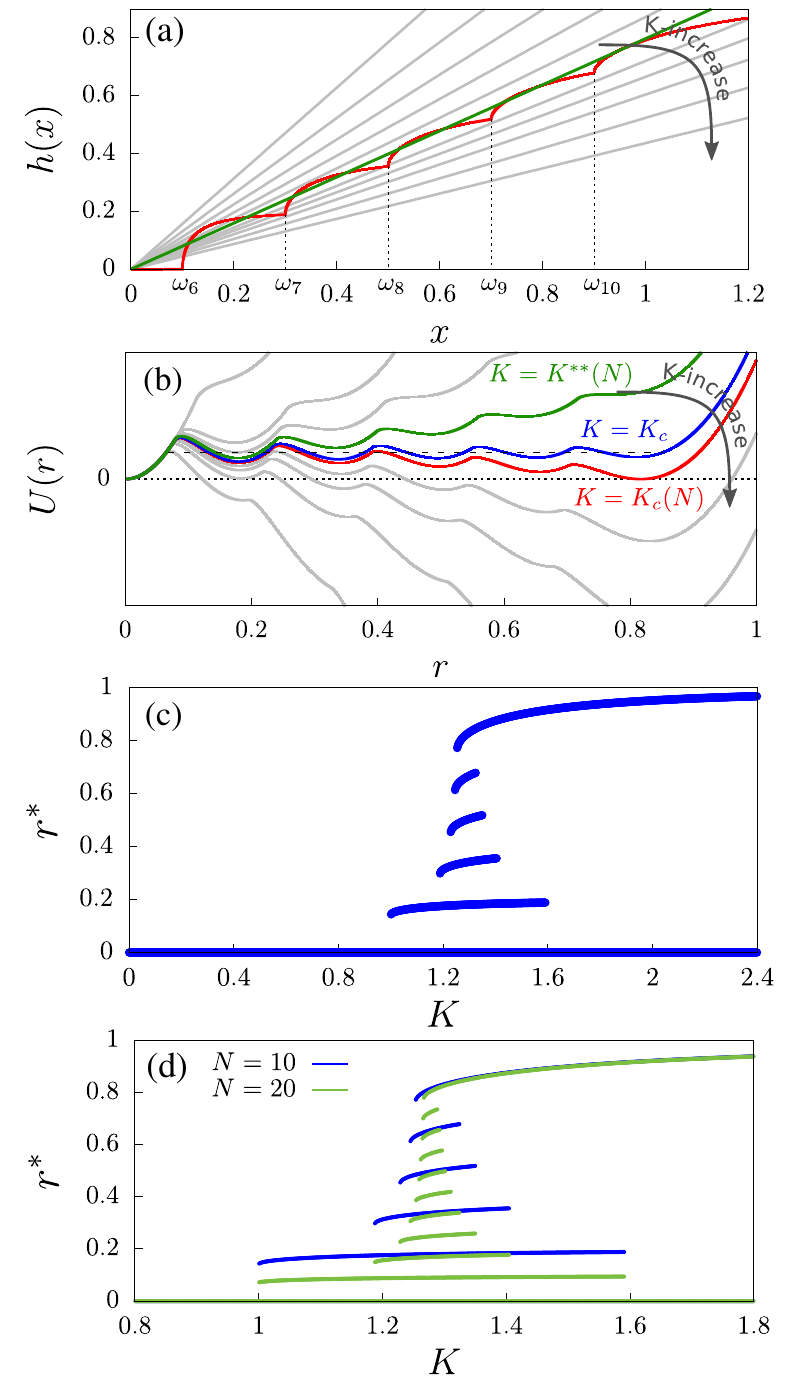}
\caption{
(a) Schematic plot of each side of Eq.~\eqref{eq:regular_finite_sce} for $N=10$ and $\gamma=1$. The r.h.s. of the equation, $h(x)$ (red line), increases abruptly at the points where $x$ is equal to each value of $\omega_i$ marked on the $x$ axis. Linear lines with various slopes indicate the l.h.s. of the equation with different values of $K$ (gray line), including $K=K^{**}(N)$ (green line). (b) Potential $U(r)$ versus $r$ at $K=K^{**}(N)$ (green), $K_c$ (blue), and $K=K_c(N)$ (red). At $K=K_c(N)$, there exist $N/2$ local minima and $U(r^*)=0$ at the largest $r^*$, representing $r_c(N)$. Thus, a global minimum occurs at $r_c(N)$. (c) Positions $r^*$ for the local minima of $U(r)$ for a given $K$ are unstable unless $U(r^*)$ is a global minimum. (d) Solutions $r^*$ of the SCE~\eqref{eq:regular_finite_sce} for two different system sizes $N=10$ and $20$.  
}
\label{fig:regular}
\end{figure}

Fig.~\ref{fig:regular}(a) illustrates the behaviors of both sides of Eq.~\eqref{eq:regular_finite_sce} along with the parameter $x$. $h(x)$ is a continuous function; however, it is not smooth in shape because the range of the summation varies with $x$. Because the slope of the l.h.s. of the equation is $1/K$, it is instructive to notice how solutions of Eq.~\eqref{eq:regular_finite_sce}, denoted as $\{r^*\}$, change as $K$ is increased by examining the crossing points of the linear line and $h(x)$. For instance, in Fig.~\ref{fig:regular}(a), when the slope $1/K$ is sufficiently large, a linear line with slope $1/K$ meets $h(x)$ only at $r^*=0$, which is a solution of the SCE. As $1/K$ is decreased, the number of solutions $r^*$ increases, and there exists $K^{**}(N)$ at which the number of solutions becomes $N$ for the first time. At this point, the SCE has a nontrivial solution in the range $x>\omega_N$, as shown in Fig.~\ref{fig:regular}(a). This solution becomes a local minimum of $U(r)$ at the largest $r^*$, denoted as $r^{**}$ which implies that all the oscillators are phase-locked. Thus, when dynamics starts from $r=1$, the system reaches to the state with the $r^{**}$ value as shown in Fig.~\ref{fig:regular}(b). When $1/K$ is decreased further and reaches $1/K_c(N)$, $U(r^{**})$ at $r^{**}$ becomes zero. This is another global minimum for finite $r$. Between these two values of $K^{**}(N)$ and $K_c(N)$, there exists the transition point $K_c(\infty)$ in the thermodynamic limit. For brevity, we denote it as $K_c$. At this $K_c$, the {\sl ad hoc} potential $U(r)$ exhibits underdamped oscillation around a plateau as depicted in Fig.~\ref{fig:regular}(b).

In Figs.~\ref{fig:regular}(c) and (d), we show the positions $r^*$ of local minima for each given $K$. For instance, when $N=10$, there exist five nonzero $r^*$ values when $K=K_c(N)$, which correspond to the positions of the five local minima in Fig.~\ref{fig:regular}(b).   


\begin{figure}
\includegraphics[width=.95\linewidth]{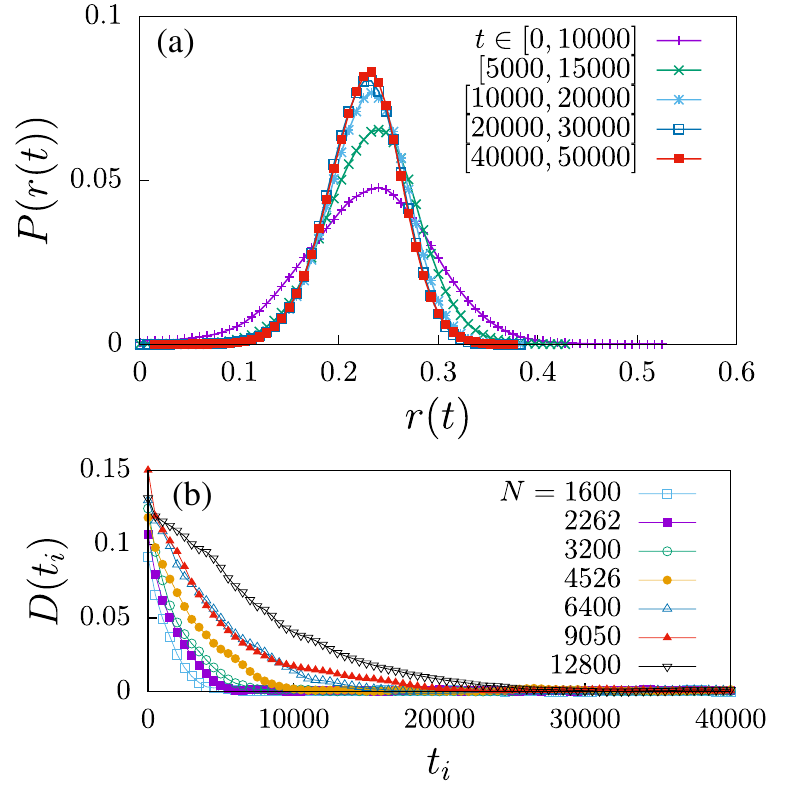}
\caption{(a) Plot of the distribution $P(r(t))$ versus $r(t)$ obtained from time intervals $t \in [0, 10000]$, $[5000, 15000]$, $[10000, 20000]$, $[20000, 30000]$, and $[40000, 50000]$ and $10^3$ realizations at $J_c(N)$. The system size is fixed as $N=6400$. (b) Plot of $D(t_i)$ versus $t_i$ for various system sizes.}
\label{fig:fss_dist_n_div}
\end{figure}

\subsection{Trapped at metastable states}
Here, we note that for the regular sampling case, the system can be more easily trapped at a longstanding metastable local minimum positioned at $r^* < r_c$. For instance, as shown in  Fig.~\ref{fig:compare_dynamics}(b), when dynamics starts from $r=1$, the system stays at $r\approx 0.8$ for a long time within the limit of our simulation time, which differs from $r\approx 0.2$ reached from an initial state with $r \sim O(N^{-1/2})$. Thus, we need more careful check if the system indeed remains at some metastable state with $r^*\ne r_c(\infty)$ as $N\to \infty$.    

We perform numerical simulations for the KM~\eqref{eq:kuramoto} with the uniform distribution of $g(\omega)$ given by Eq.~\eqref{eq:uniform_dist} at a fixed $K_c=4\gamma/\pi$. The system size is controlled. We remind that at $K_c$, the potential $U(r)$ exhibits underdamped oscillation around a plateau, whereas at $K_c(N)$, the potential $U(r)$ is slanted. We first assign a random set of initial phases $\{\theta_i(0)\}$ ($i=1,\dots,N$) and trace the order parameter as a function of time for $10^3$ realizations. For better statistics, we take time intervals specified in the legend of  Fig.~\ref{fig:fss_dist_n_div}(a). Each of these intervals contains $10^4$ times. Taking all order parameter values in each given time interval, the distribution of the order parameter $P(r(t))$ is constructed as shown in Fig.~\ref{fig:fss_dist_n_div}(a). Whereas in early time intervals, the order parameters are distributed in broad range of $r$, as time goes on, the distribution becomes narrower; the mean value is shift; and it finally approaches to a stationary distribution, being insensitive to when the interval is taken. To check the stability of the distribution function, we use the so-called Kullback-Leibler (KL) divergence, in which measure $D$ is introduced as 
\begin{align}
D \equiv \int a(r) \ln \left( \frac{a(r)}{b(r)} \right)dr + \int b(r) \ln \left( \frac{b(r)}{a(r)} \right)dr. \label{eq:KL}
\end{align}
This measure indicates to what extent two distributions $a(r)$ and $b(r)$ differs from each other. When the two distributions are exactly the same, $D=0$. To check the KL divergence for $P(r(t))$, we take the $P(r(t))$ obtained from the latest time interval $t \in [9.9\times 10^4, 10^5]$ as $a(r)$ and the distribution at different time interval $[t_i, t_i+1000]$ as $b(r)$. Then, the dependence of $D$ on $t_i$ is calculated with increasing $t_i$. Since the distribution $P(r)$ converges to a certain form as illustrated in Fig.~\ref{fig:fss_dist_n_div}(a), we expect that $D$ gradually decreases and approaches to zero. Indeed, $D(t_i)$ behaves as shown in Fig.~\ref{fig:fss_dist_n_div}(b). Moreover, we trace $D(t_i)$ values as a function of $t_i$ for different system sizes $N$, finding that the saturation time becomes longer as the system size is increased. Based on these results, we conclude that the distribution $P(r)$ for $N \le 12800$ is in steady state at the time $t=10^5$. 

\begin{figure}
	\includegraphics[width=.95\linewidth]{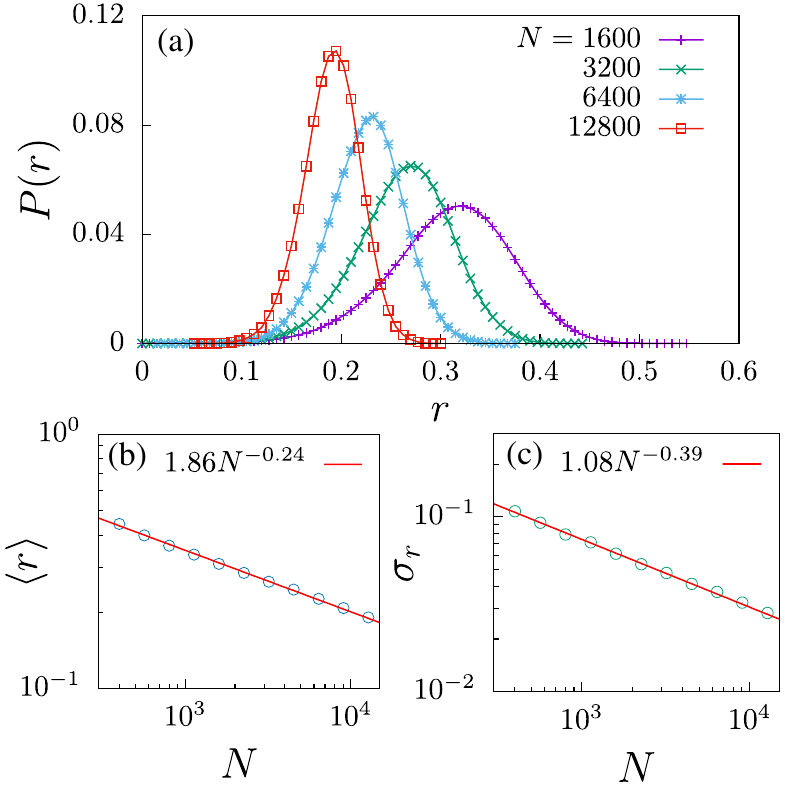}
	\caption{(a) Plot of the distribution $P(r)$ versus the order parameter value $r$. Data points are obtained from different $10^6$ time steps in steady state and $10^3$ samples for various system sizes $N$ at $K_c(\infty)$. (b) Plot of $\langle r \rangle$ versus $N$. The straight line is a guideline with slope $-0.24$. (c) Plot of the standard deviation of $P(r)$ versus $N$. The straight line is a guideline with slope $-0.39$.}
	\label{fig:fss_r_dist}
\end{figure}

\begin{figure}[h!]
	\includegraphics[width=.95\linewidth]{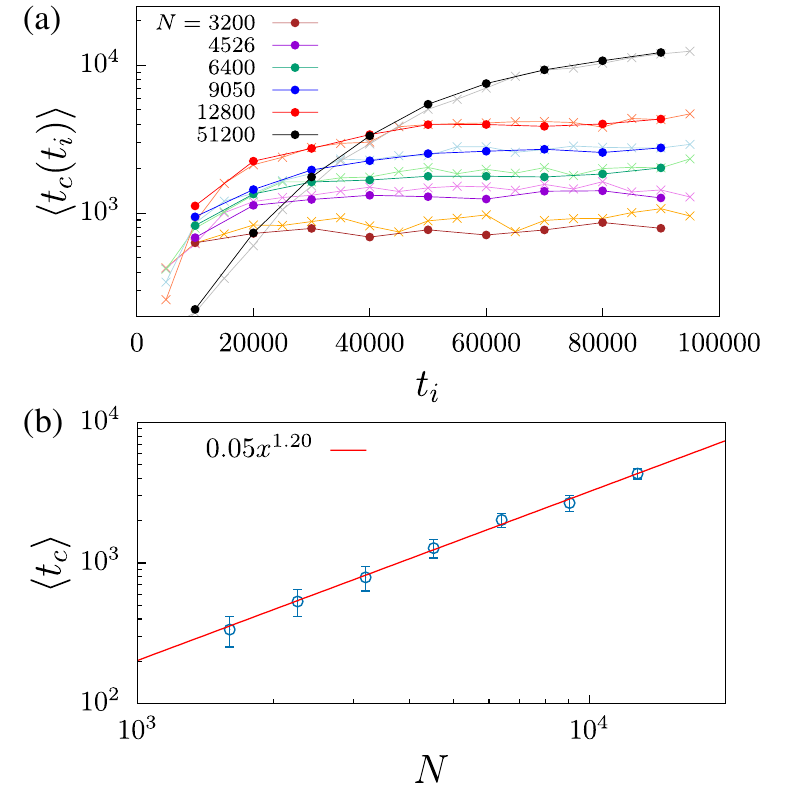}
	\caption{(a) Plot of $\langle t_c(t_i) \rangle$ for various system sizes. The data points are averaged over $10^3$ samples. The order parameter $r(t)$ was measured until time $t=10^5$ in steps of $\Delta t=5000$ (light, cross) and $10000$(dark, circle) to  evaluate $t_c$. The obtained values of $\langle t_c \rangle$ seems to be saturated for $N \le 12800$, while they are not for the case of $N=51200$. (b) Plot of the estimated values of $\langle t_c \rangle$ versus $N$ up to 12800. The straight line is a guideline with estimated slope $1.20$.}
	\label{fig:fss_tc}
\end{figure}

We examine the distribution $P(r)$ for different system sizes $N$ in steady states. As the system size $N$ is increased, the peak position of $P(r)$ moves to the left and the width becomes narrower, as shown in Fig.~\ref{fig:fss_r_dist}(a). By measuring the mean values  $\langle r \rangle$ of $P(r)$ and the standard deviation $\sigma_r$ for various system sizes, we obtain power-law decays as $\langle r \rangle \sim N^{-0.24}$ and $\sigma_r \sim N^{-0.39}$, as shown in Figs.~\ref{fig:fss_r_dist}(b) and~\ref{fig:fss_r_dist}(c), respectively. These power-law behaviors suggest that the system stays at $r=0$ in the limit $N\to \infty$, which is in agreement with the previous result in Sec.~\ref{subsec:uniform}. Moreover, this result may explain the reason for the discrepancy of the steady states reached from different initial configurations for the regular sampling case shown in Fig.~\ref{fig:compare_dynamics}(b). 

Finally, we estimate a characteristic time $\langle t_c \rangle$, beyond which the order parameter reaches a steady state. We perform simulations up to $t=10^5$ for the system size $N \le 51200$ in the following way. First, we take time intervals $[t_i, t_i+\Delta t]$, where $t_i$ is taken as the dotted ones in Fig.~\ref{fig:fss_tc}(a) and $\Delta t$ is taken appropriately as represented in the caption of Fig.~\ref{fig:fss_tc}. Second, the order parameter is averaged over each time interval, which is denoted as $\bar r (t_i)$. Next, we determine the characteristic time $t_c(t_i)$ at which $r(t)$ becomes larger than $\bar r (t_i)$ for the first time. We repeat this process until $t_i+\Delta t=10^5$. Next, $t_c(t_i)$ are averaged over $10^3$ realizations, and the resulting mean is denoted as $\langle t_c(t_i) \rangle$. 
Fig.~\ref{fig:fss_tc}(a) shows that $\langle t_c(t_i) \rangle$ seems to be saturated to a constant value (denoted as $\langle t_c \rangle$) as $t_i$ is increased for $N \le 12800$. However, when $N=51200$, the simulation time $10^5$ seems to be insufficient, and longer simulation time is required. Finally, we check the characteristic time $\langle t_c \rangle$ as a function of $N$. Fig.~\ref{fig:fss_tc}(b) shows that $\langle t_c \rangle$ exhibits power-law behavior with respect to $N$ as $\langle t_c \rangle \sim N^{1.2}$. Thus, the dynamic exponent for the system size $N$ is estimated to be $\bar z\approx 1.2$.


\section{Summary and Discussion}\label{sec:summary}
We reconsidered the hybrid synchronization transitions arising in the KM by constructing an {\sl ad hoc} potential analogous to the Landau free energy conventionally used in thermal equilibrium systems. In particular, we considered KEs with several different types of natural frequency distributions which generate hybrid synchronization transitions. From the SCEs of the KMs, we constructed {\sl ad hoc} potentials and showed that the {\sl ad hoc} potential in the thermodynamic limit satisfies the criterion of the Landau theory for an HPT established for thermal systems~\cite{at}. 

For finite systems, the landscape of the {\sl ad hoc} potential contains a finite number of local minima created by the natural frequencies of entrained oscillators. The barrier height between two consecutive local minima becomes lower as $N$ is increased as we compare Fig.~\ref{fig:uniform_potential}(a) to Fig.~\ref{fig:regular}(b). The energy barrier between them near $r=0$ is overcome by fluctuations of the coherence due to drifting oscillators. If we can ignore the correlation effect between synchronized and drift oscillators, then the strength of these fluctuations would be proportional to $\sqrt{\langle N_d\rangle}/N$, where $N_d$ is the number of drifting oscillators given as $\langle N_d \rangle=(1-r/r_c)N$. Thus, the strength is weakened as $N$ is increased and $r$ approaches $r_c$. When the fluctuations become too small to overcome the barrier height between nearby local minima, the system is trapped at a metastable position, and its average is the mean value $\langle r \rangle$ of the distribution $P(r)$. As we observed in numerical simulations, the mean $\langle r \rangle$ is reduced as $N$ is increased.  

Furthermore, the landscape provides an intuitive understanding of the dependence on the initial phases for regularly and randomly chosen sets of natural frequencies. We applied the  proposed methodology to the Kuramoto systems with various sets of natural frequencies and coupling strengths for diverse types of synchronization transitions such as hybrid, second-order, and first-order transitions. Consequently, this approach could be useful for determining transition types of synchronizations and understanding transition properties for other Kuramoto-type models. \\

\begin{acknowledgments}
This research was supported by the National Research Foundation of Korea (NRF) through grant no. NRF-2014R1A3A2069005 (BK). The authors thank Professors S. Dorogovtsev and J.F.F. Mendes for helpful discussions. 
\end{acknowledgments}

\end{document}